# Oblique Bragg-Doppler scattering at relativistic grating created by backward Raman amplification in hydrogen

Wenhong Lai[1,2,3], Jiapeng Huang[1,*], Haozhe Guo[1], and Philip St.J. Russell[1]

[1]Russell Centre for Advanced Lightwave Science, Shanghai Institute of Optics and Fine Mechanics and Hangzhou Institute of Optics and Fine Mechanics, Chinese Academy of Sciences (CAS), Shanghai, China
[2]University of Chinese Academy of Sciences, Beijing, China
[3]Zhejiang Key Laboratory of Microstructured Specialty Optical Fiber, Hangzhou Institute of Optics and Fine Mechanics, Hangzhou, China

*Corresponding author: jiapenghuang@siom.ac.cn

**Abstract**

We report the first observation of relativistic "Bragg-Doppler" reflection, which occurs when a light beam is incident on a fine-period phase grating moving at close to the speed of light. The grating is created in hydrogen-filled hollow-core fibre by seeded backward stimulated Raman scattering, which creates an intense coherence wave of molecular vibrations at 125 THz, with wavelength 299 nm, moving at ~$c/8$. When this grating is obliquely probed at a generalized Bragg angle, a reflected beam emerges, Doppler-shifted by 125 THz. The effect is also uni-directional, as phase-matching is strongly violated when the probe beam is reversed, so may be viewed as an ultra-high frequency Bragg-cell frequency-shifter. The wide transparency window of hydrogen allows frequency shifting of radiation from the visible to the vacuum UV, simply by tuning the incident angle. The results open up new opportunities for spatio-temporal studies of coherence wave dynamics, frequency conversion in difficult-to-access spectral regions such as the deep and vacuum UV, and quantum-state-preserving frequency conversion of single photons.

## 1 Introduction

Hollow-core photonic crystal fibre (HC-PCF) is ideal for exploring gas-based nonlinear optics in a diffraction-free environment, while offering excellent mode quality, pressure-tunable dispersion, and very low transmission loss over a broad spectral window[1,2]. This has made possible efficient stimulated Raman scattering (SRS) in hydrogen at unprecedentedly low threshold powers[3,4], and light generation from the UV to the mid-IR[5,6]. SRS itself is a process in which a fast-moving refractive index wave, created by excitation of coherent molecular vibrations, causes phase-matched frequency down-conversion of a pump signal at frequency $\omega_P/2\pi$ to a Stokes signal at $(\omega_P - \Omega)/2\pi$, where $\Omega/2\pi$ is the Raman transition frequency. In contrast to forward SRS, backward SRS (BSRS) strongly dominates over forward SRS above a certain threshold power, as observed in gas cells[7–9] and gas-filled HC-PCF[10]. In contrast to the forward case, when the coherence wave is weak and distributed over the entire fibre length, in BSRS it grows exponentially with pump power while its effective interaction length shrinks[7]. Indeed, almost all the incident pump energy is converted to a backward Stokes signal over interaction lengths as short as a fraction of a mm.

As the coherence wave (Cw) is a strong fine-period refractive index grating, with modulation depth as high as ~$10^{-4}$ in our experiments, it may be used within its coherence lifetime (~0.5 ns at 15 bar pressure) to Bragg scatter and Doppler shift an obliquely incident signal. Here we report the first observation of this "Bragg-Doppler scattering", close to the input face of a hydrogen-filled HC-PCF (Fig. 1a). Pump pulses at $\lambda_P = 532$ nm (2 ns, 25 µJ) are launched into the fibre, and the vibrational transition (125 THz) is driven by backward-propagating seed pulses at the Stokes wavelength ($\lambda_S = 683$ nm, 0.1 µJ). This results in the creation of a coherence wave with wavelength $\Lambda_C = \lambda_P\lambda_S/(\lambda_S + \lambda_P) = 299$ nm, travelling into the fibre at a phase velocity ~$c/8$, growing exponentially in strength towards the input face of the fibre, and extending jet-like ~500 µm into free space. It may be used to Doppler-shift obliquely incident light by $\pm\Omega/2\pi = 125$ THz at wavelengths from the visible to the deep vacuum ultraviolet, simply by adjusting the probe angle to satisfy the Bragg-Doppler condition. As well as offering broad-band frequency conversion, even in difficult-to-access spectral regions such as the vacuum UV, the technique provides a means, for the first time to our knowledge, of directly probing the transient spatio-temporal evolution and polarization dependence of the Raman coherence in BSRS.

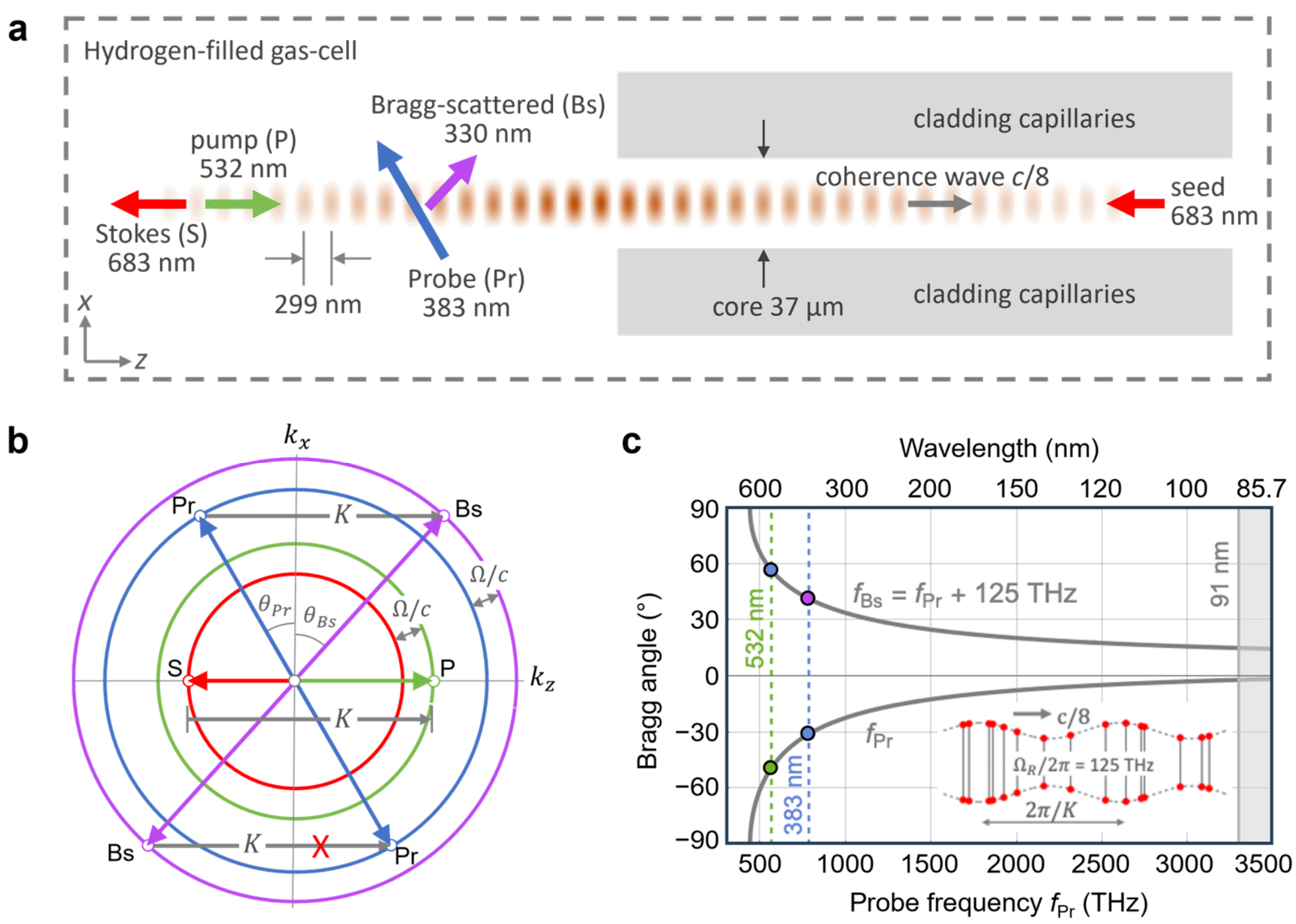

**Fig. 1: a**, Illustration of the experiment (not to scale). A Raman coherence wave, created in a hydrogen-filled HC-PCF by pumping (P) with 532 nm pulses and backward seeding with Stokes (S) light at 683 nm, extends jet-like ~0.5 mm outside the fibre end-face, where in the illustration it is obliquely probed with a separate 383 nm beam at the Bragg angle, resulting in a Doppler up-shifted signal at 330 nm. **b**, Wavevector diagram used to find the phase-matching condition for oblique Bragg scattering and Doppler up-conversion of an arbitrary-wavelength probe signal (Pr) by the coherence wave. See text for more details. **c**, Incident and scattered Bragg angles, measured clockwise from the *x*-axis, for phase-matched frequency up-conversion by $\Omega/2\pi =$ 125 THz of a probe signal ($f_{Bs} = f_{Pr} + \Omega/2\pi$). The absorption edge of hydrogen is at ~91 nm. The dashed lines correspond to a 383 nm probe (Fig. 1(a&b)) and the 532 nm probe used in the experiment. The inset shows a classical picture of a coherence wave (see text).

# 2 Theoretical description

A wavevector diagram (Fig. 1b) provides a convenient means of finding the phase-matching condition for Bragg-Doppler scattering of an obliquely incident arbitrary-wavelength probe signal (Pr) by the coherence wave. The coloured circles are loci $k_x^2 + k_z^2 = (\omega_i n/c)^2$ of the allowed wavevectors in the *xz* plane at frequency $f_i = \omega_i/2\pi$, with refractive index $n$ taken to be unity for all the waves (although this is not exactly true for the guided modes, the error is very small). Figure 1b illustrates the case when a coherence wave with momentum $K = (2\omega_P - \Omega)/c$, created by pumping (P) with 532 nm pulses (frequency $f_P$) and backward seeding with Stokes (S) light at 683 nm, is probed at 383.1 nm, yielding a Bragg scattered signal (Bs) at 330.4 nm. Note that the diagram may also be used for frequency down-conversion if the labels Bs and Pr are interchanged. The incident and scattered Bragg angles $\theta_{Pr}$ and $\theta_{Bs}$ are unequal because of the Doppler frequency shift:

$$\sin\theta_{Pr} = \frac{f_{Pr}f_R - 2f_P(f_P - f_R)}{f_{Pr}(2f_P - f_R)}, \qquad \frac{\sin\theta_{Pr}}{\sin\theta_{Bs}} = \frac{f_{Bs}(f_{Pr}f_R - 2f_P(f_P - f_R))}{f_{Pr}(f_{Bs}f_R + 2f_P(f_P - f_R))} \tag{1}$$

where $f_P = 563.9$ THz (532 nm) and $f_S = 439.2$ THz (683 nm) are the pump and seed frequencies used to write the coherence wave, $f_{Pr}$ is the frequency of the probe light, $f_{Bs} = f_{Pr} + f_R$ the frequency of the Bragg-scattered light, and $f_R = \Omega/2\pi$ is the Raman frequency. The calculated Bragg angles, measured clockwise from the *x*-axis, for probe wavelengths from 683 to 91 nm (the absorption edge of hydrogen), are plotted in Fig. 1c. For the case illustrated in Fig. 1b, $f_{Pr} = 783$ THz (383 nm) and we find $\theta_{Pr} = -30.4°$ and $\theta_{Bs} = +41.9°$ and a Bragg scattered wavelength of 330 nm. In the experiment $f_{Pr} = f_P$, yielding $\theta_{Pr} = \sin^{-1}(-(2f_P - 3f_R)/(2f_P - f_R)) = -48.6°$, $\theta_{Bs} = +57.2°$, and a Bragg scattered wavelength of 435 nm. By scanning the probe along the *z* direction, the spatio-temporal behaviour of the coherence wave, and the polarization-dependence of the scattering, can be investigated.

Viewed classically, the coherence wave (inset in Fig. 1c) is a population of vibrating molecules whose relative phase varies sinusoidally in space. Although on the timescale of the interaction the excited molecules remain in the same (random) locations (in reality their orientations are also random), their oscillations are phased so as to create a sinusoidal coherence wave (dashed envelope) that travels at $\sim c/8$ in the +*z* direction.

The conversion efficiency between probe and Bragg scattered signals can be estimated using coupled mode theory[11], modified to include the Doppler shift at a rapidly moving grating (see section 1 of supplementary information (SI)). The power conversion efficiency to the Bragg scattered signal takes the form, assuming negligible probe depletion:

$$\eta_{Bs} = \frac{f_{Bs}^2 \cos\theta_{Bs}}{f_{Pr}^2 \cos\theta_{Pr}} \frac{\sin^2\left(\mathrm{h}\sqrt{\kappa_{Pr}\kappa_{Bs} + \vartheta^2/4}\right)}{(1 + \vartheta^2/(4\kappa_{Pr}\kappa_{Bs}))} \tag{2}$$

where $h$ is the effective width of the coherence wave, $\kappa_j = n_R \pi f_j/(2c\cos\theta_j)$ are coupling constants, $n_R$ is the effective index modulation of the coherence wave, and $\vartheta = (\omega_{Bs}\cos\theta_{Bs} - \omega_{Pr}\cos\theta_{Pr})/c$ is the dephasing parameter, which for small deviations from the incident Bragg angle $\theta_{Pr}^0$ can be approximated as:

$$\vartheta = \frac{-4\pi f_{Pr}/c}{\sqrt{(2f_{Pr} + f_R)^2/f_K^2 - 1}}\Delta\theta_{Pr}, \qquad \Delta\theta_{Pr} \ll \pi \tag{3}$$

where $\Delta\theta_{Pr} = (\theta_{Pr} - \theta_{Pr}^0)$ and the approximation $(f_R/f_K)^2 \ll 1$ has been used (see section 1 of SI). Note that the coherence wave will have a smooth bell-shaped transverse profile, resulting in apodization that will flatten the side-lobes predicted by Eq.(2).

# 3 Experiment

The experimental set-up is sketched in Fig. 2a. Two ~1 m lengths of HC-PCF are connected at each end to gas-cells that are first evacuated and then filled with hydrogen to 6 bar for seed generation and 15 bar for BSRS. One of the gas-cells is equipped with glass windows on two sides, to allow oblique probing of the coherence wave during BSRS. The HC-PCF was of single-ring design (scanning electron micrograph in Fig. 2b), with six capillaries (inner diameter ~17 μm, capillary wall thickness 370 nm) surrounding a hollow core diameter of ~37 μm. The transmission loss of the HC-PCF was less than 0.1 dB/m over the spectral range ~450 nm to ~720 nm. Light from a 532 nm pump laser, delivering ~2 ns pulses at a repetition rate of 10 kHz, was split in two parts and launched into the two HC-PCFs (Fig. 2a). A Stokes signal (683 nm) was generated by forward SBS in the first fibre and, after spectral filtering to remove residual pump and unwanted anti-Stokes light (435 nm), was used to seed BSRS in the second. The HC-PCF acts as a modal filter, ensuring that the seed signal is in a pure $LP_{01}$-like mode.

Strong amplification of the backward seed was observed at 25 μJ pump and 0.1 μJ seed energy. A small fraction of the 532 nm pump laser power was diverted at a beam-splitter and focused obliquely, through the gas-cell window, into the coherence wave. Two adjustable optical delay lines were used to synchronize the three pulses so as to maximize the Bragg-scattered signal. For probe angle $\theta_{Pr} \approx -49°$ a weak purple-coloured spot was visible to naked eye on a screen, corresponding to a Bragg-scattered beam at $\theta_{Bs} \approx +57°$. Both angles are in good agreement with the predictions of theory (Eq. (1)).

The coherence wave, measured ~50 μm from the fibre end-face, reached maximum strength when the pump and seed pulses had energies 25 μJ and 0.1 μJ, with peaks coinciding in time and space (in the absence of any SRS) at a point ~10 cm inside the fibre: too close to the fibre end and the seed amplification is insufficient, deeper inside the fibre and the coherence wave does not reach the fibre end-face. Incidentally, we were able to detect a Bragg-scattered signal when probing the coherence wave through the fibre cladding, but scattering in the cladding structure made the results hard to interpret precisely.

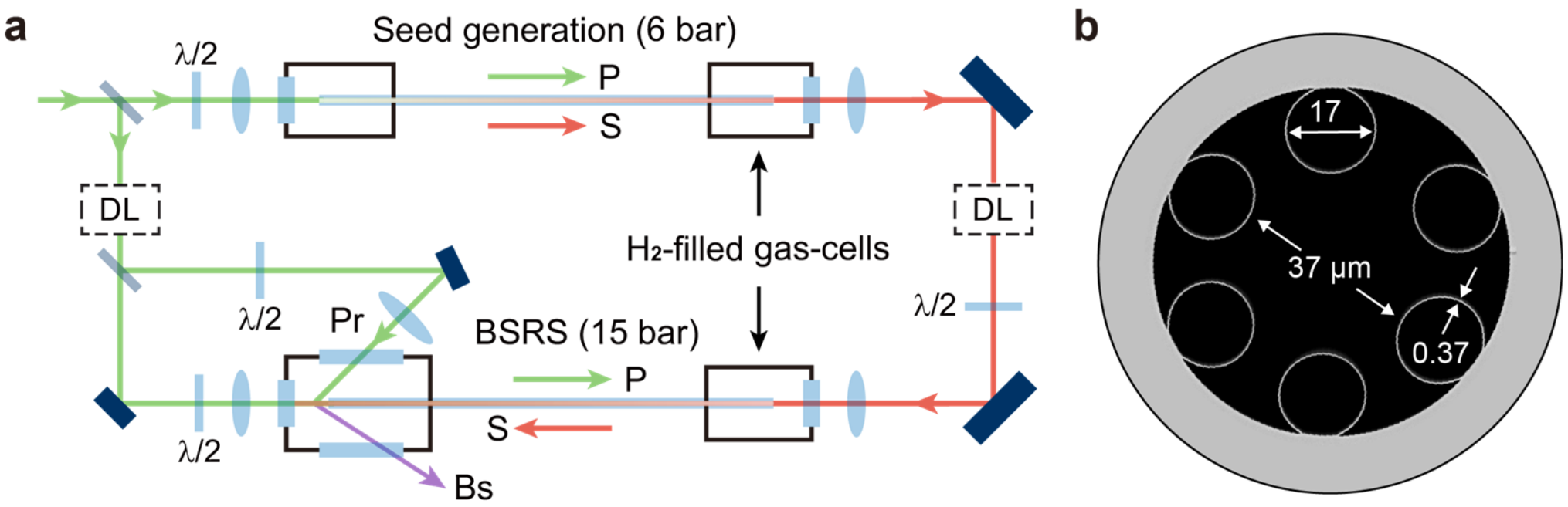

**Fig. 2:** **a**, Experimental set-up. The light blue rectangles are glass windows, and DL = variable delay line. See text for more details. **b**, Scanning electron micrograph of the HC-PCF structure (dimensions in μm).

The intensity profile of the Bragg-scattered beam was imaged in the near-field close to the coherence wave (Fig. 3a). It is elliptical in shape, the short axis (~11 μm) corresponding approximately to the width of the coherence wave, and the long axis to the width of the probe beam, i.e., $w_{Bs} = w_{Pr} \cos\theta_{Bs} / \cos\theta_{Pr}$ (Fig. 3b). Taking account of the overlap between the probe beam and the coherence wave, the peak measured conversion efficiency from probe to Bragg-scattered signal was $3.9 \times 10^{-5}$. By scanning the probe beam along the $z$-axis outside the fibre end-face and monitoring the Bragg scattered signal we were able to measure the profile of the coherence wave amplitude (Fig. 3c). Reasonable agreement with theory, including the effect of beam diffraction on pump/seed overlap, is obtained, showing a decay length of ~0.5 mm. The angle-dependence of the scattered signal is plotted in Fig. 3d, along with a fit to Eq.(2) with $h = 13$ μm (slightly larger than the beam width) and $n_R = 0.95 \times 10^{-4}$. The agreement is good. The absence of side-lobes in the experimental data we attribute to the smooth Gaussian-like transverse profile of the coherence wave.

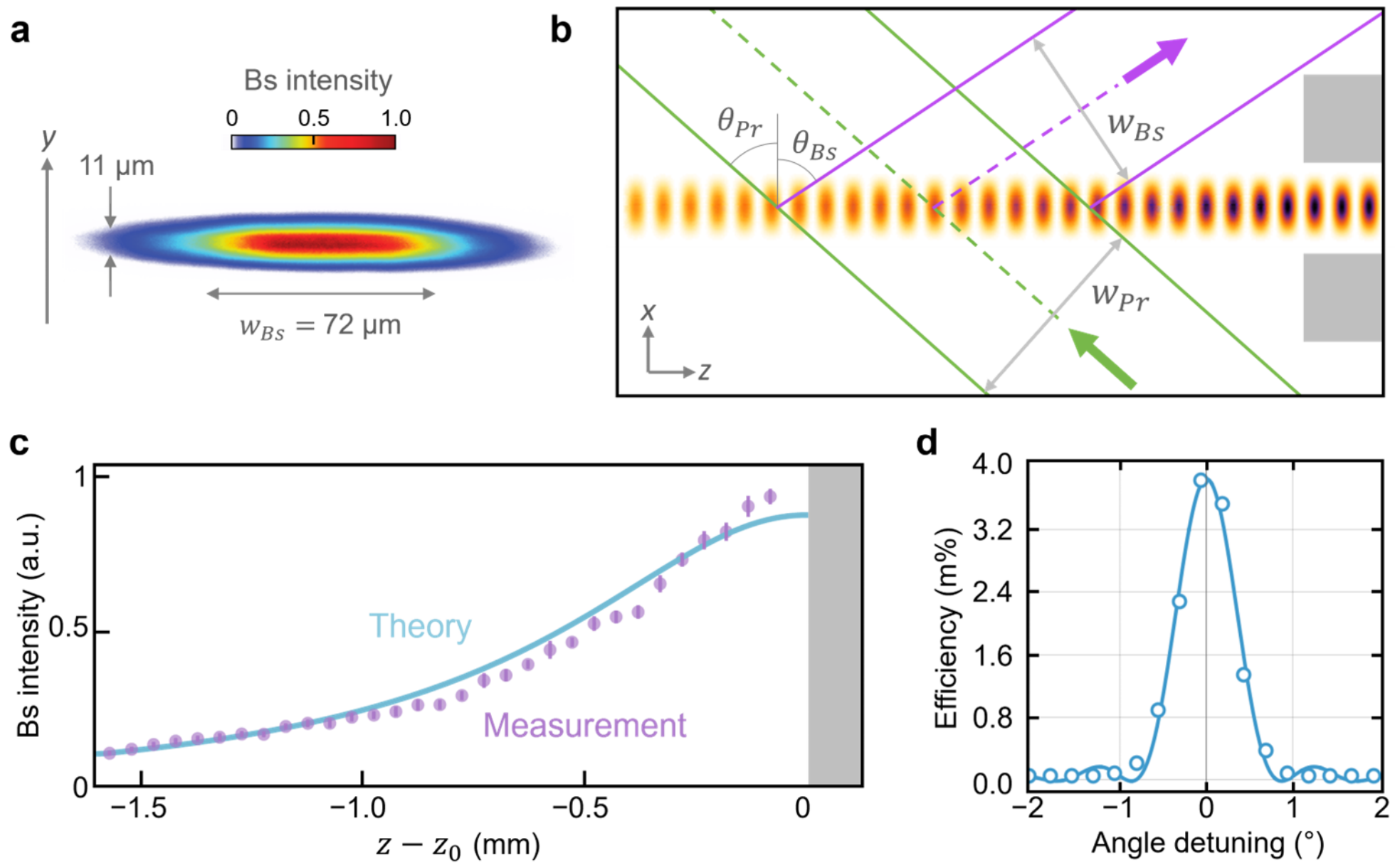

**Fig. 3:** **a**, Near-field image of the Bragg-scattered beam profile. Its width in the $y$-direction (11 μm) corresponds roughly to the width of the coherence wave. **b**, Geometry of the set-up for measuring the Bragg-scattered signal (435 nm) for a probe

beam at 532 nm (schematic only). The fibre end-face is at $z = z_0$. In the experiment the beam widths and angles are $w_{Pr} =$ 88 µm, $|\theta_{Pr}| = 48.6°$, $w_{Bs} = 72$ µm, $|\theta_{Bs}| = 57.2°$. **c**, Theoretical (blue solid line) and experimental (purple dots) signal intensity as a function of distance from the fibre tip ($z = z_0$). The decay length is ~0.5 mm. **d**, Conversion efficiency from probe to Bragg scattered signal as a function of probe angle, based on the fraction of the probe light that intersects with the coherence wave. The open circles are experimental values, and the curve is a fit to Eq.(2) with parameters $h = 13$ µm and $n_R = 0.95 \times 10^{-4}$.

The spatio-temporal behaviour of pump, Stokes and coherence wave was modelled using a previously reported numerical technique[10]. Under optimized conditions in the modelling, the transmitted pump wave is almost completely depleted, and the coherence wave has an overall length of ~2 mm, 0.5 mm of it extending outside the fibre end-face, in close agreement with measurements (Fig. 3c). In the experiments, the Bragg scattering at $z = -50$ µm is strongest for a seed energy of 0.4 µJ (Fig. 4a). As the seed energy is increased first to 0.07 µJ and then to 0.4 µJ, keeping the pump power constant, the measured coherence wave and seed strengths at $z = -50$ µm increase while their temporal profiles exhibit more and stronger oscillations, as energy is exchanged periodically between pump, seed and coherence wave (Fig. 4b). Although similar "accordion-like" [12] behaviour has been reported previously in forward SRS[13,14], this is the first time such oscillations have been directly observed and measured in BSRS.

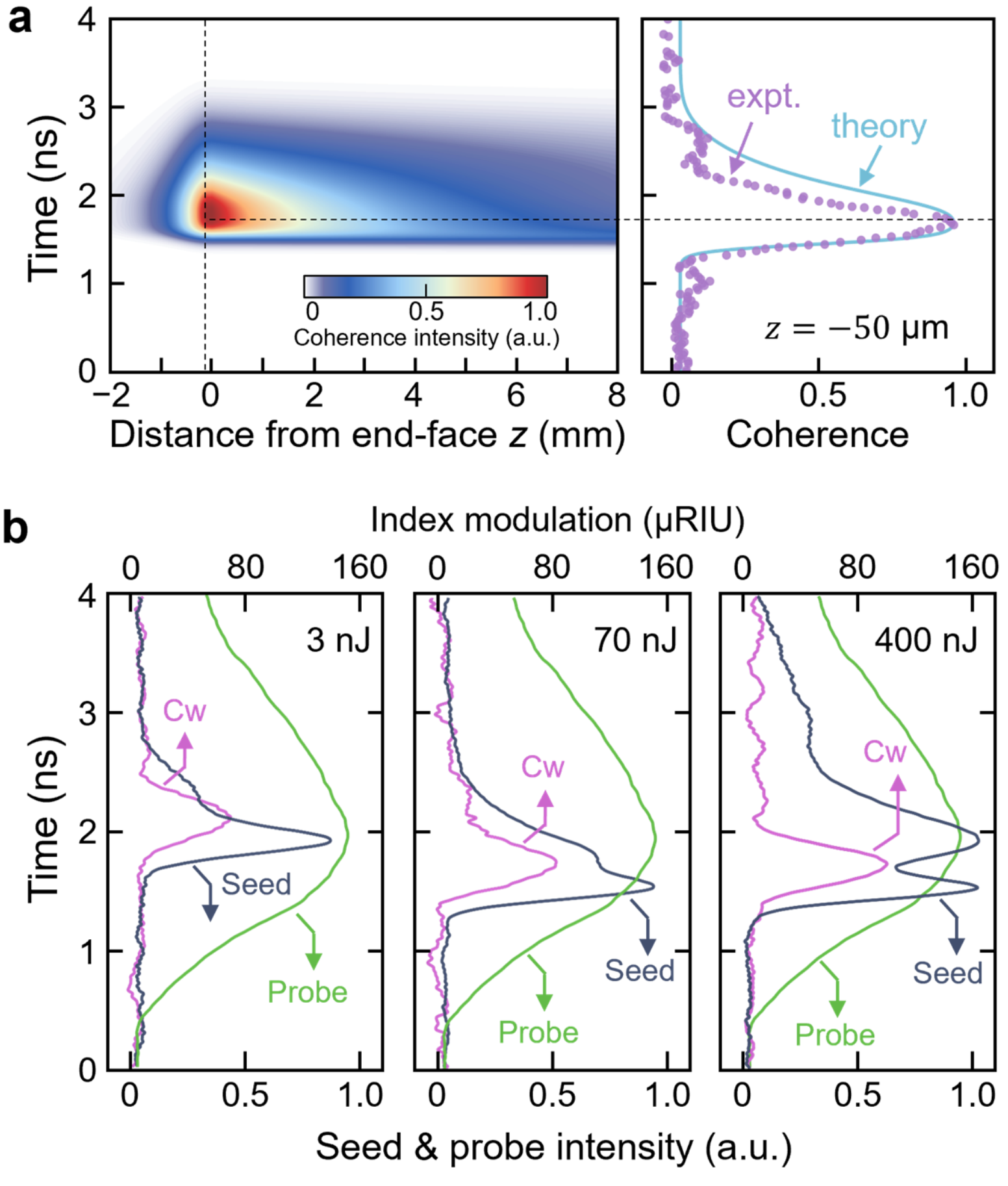

**Fig. 4: a**, Left: Simulated spatio-temporal evolution of the coherence wave (Cw) amplitude for 25 µJ pump and 0.1 µJ seed, using the numerical approach described in[10] (see section 1 & 2 of SI). Right: Measured (red line) and simulated (blue line) of the temporal profiles of the coherence wave amplitude at $z = -50$ µm. In both plots, 1.0 corresponds to a coherence wave amplitude of ~100 µRIU. **b**, Measured time dependence, at $z = -50$ µm, of the probe pulse (green), the amplified seed pulse (black), and coherence wave strength (Cw, magenta) for seed energies 0.003 µJ, 0.07 µJ and 0.4 µJ. The oscillatory behaviour, a sign of "accordion" effects[12], is more pronounced at higher seed energies.

The Bragg-scattered power depends on the polarization states of pump, seed and probe pulses. Since in the experiment the pump and seed are co-linearly polarized, the maximum vibrational Raman polarization follows their electric field, which points at angle $\phi$ to the *y*-axis in the *xy*-plane. To estimate the polarization-dependence of the Bragg scattered power, we first transform the probe electric field into the laboratory frame, scatter it off the Raman polarization, and then resolve the resulting field into the Bragg-scattered frame. As illustrated in Fig. 5a, the Bragg-scattered power can be expressed mathematically as:

$$p_{Bs} \propto \left| \hat{\mathbf{u}}_{Bs} \cdot [\mathbf{R}]_{\vec{y}}^{\pi/2-\theta_{Bs}} \cdot [\mathbf{R}]_{\vec{z}}^{\phi} \cdot [\boldsymbol{\varepsilon}_R] \cdot [\mathbf{R}]_{\vec{z}}^{-\phi} \cdot [\mathbf{R}]_{\vec{y}}^{-\pi/2-\theta_{Pr}} \cdot \hat{\mathbf{u}}_{Pr}(\theta) e_{Pr} \right|^2 \tag{4}$$

where $e_{Pr}$ is the magnitude of the probe field, $\hat{\mathbf{u}}_{Pr} = (\sin\theta, \cos\theta, 0)$ and $\hat{\mathbf{u}}_{Bs} = (\sin\psi, \cos\psi, 0)$ are unit vectors pointing along the electric field of the Pr and Bs signals in their local frames, $[\boldsymbol{\varepsilon}_R] = \mathrm{diag}(\varepsilon_R/3, \varepsilon_R, \varepsilon_R/3)$ is a dielectric tensor representing the Raman polarization, and $[\mathbf{R}]_{\vec{q}}^{\xi}$ is a Rodrigues matrix that resolves a vector into a new coordinate system rotated by angle $\xi$ about an axis defined by unit vector $\vec{q}$ (see section 3 of SI). In the experiment we measured the Bragg-scattered power as a function of probe polarization angle $\theta$ for $\phi$ = 0°, 45° and 90°. The results show good agreement with theory (Fig. 5b).

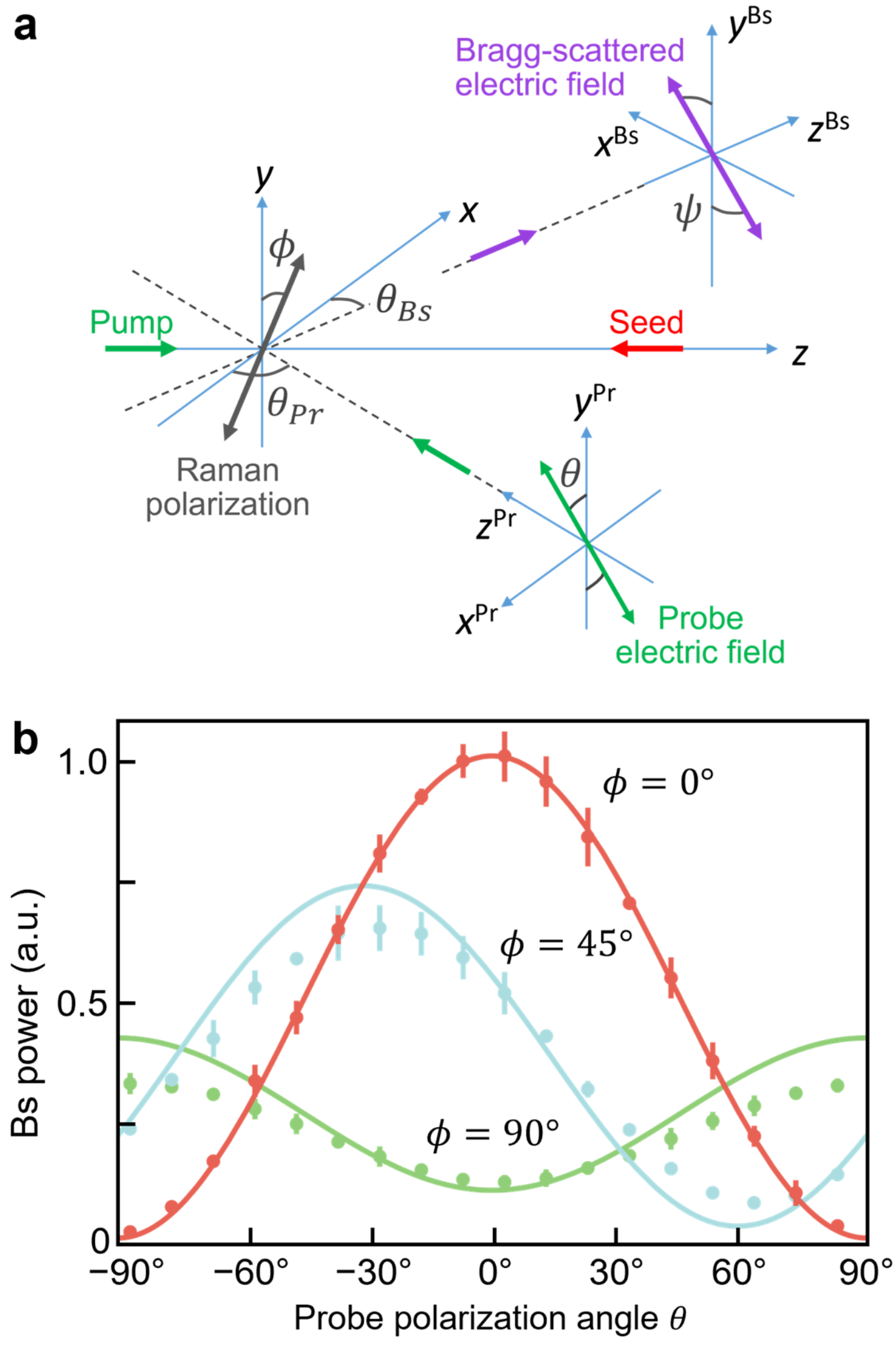

**Fig. 5:** **a**, Diagram for three-dimensional vectorial analysis of oblique Bragg scattering. The laboratory frame is *xyz*, the pump and seed beams travel along the ±*z*-axis, and the direction of the Raman polarization (in the *xy*-plane at angle $\phi$ to the *y*-axis) is set by the co-linearly polarized electric fields of pump and seed. The probe and Bragg scattered fields are linearly polarized as shown, within their own coordinate systems. **b**, Measured (dots) and calculated (solid lines) power in the Bragg-scattered signal for different pump and probe polarization states, normalized to the value when pump, seed and probe are co-polarized perpendicular to the *xz*-plane, i.e., $\phi = \theta = 0$.

## 4 Discussion & conclusion

We note that the efficiency of Bragg-Doppler scattering can be enhanced by using a HC-PCF with a larger core, increasing the pump energy and optimizing the seed pulse energy and timing to maximize the coherence wave strength outside the fibre. Since only a small fraction of the coherently excited molecules is used for up-conversion, the scattering may be regarded as a linear process[15]. By tuning the probe angle, therefore, it can be used for quantum-state-preserving frequency up- or down-conversion of single photons over a very wide range of probe frequencies, even, since hydrogen remains transparent down to ~90 nm, deep into the vacuum UV (Fig. 1c). An additional interesting feature of Bragg-Doppler scattering is that it is uni-directional, i.e., if the probe beam is reversed in direction, there is no Bragg reflection (Fig. 1b).

In conclusion, this is the first report of Bragg-Doppler scattering of an obliquely incident probe beam by the intense Raman coherence wave that forms during backward SRS in hydrogen-filled HC-PCF. A strong fine-period diffraction grating moving at relativistic speeds is created within the fibre and extends jet-like ~500 μm into free space, where it can be probed obliquely at a different frequency. The technique permits full-field spatio-temporal measurements of the complex coherence wave dynamics, including its polarization dependence. The results open up new opportunities for frequency conversion in difficult-to-access spectral regions such as the deep and vacuum UV.

## 5 Author contributions

JH and WL designed the experimental setup and carried out the measurements, assisted by HG. PR proposed the idea and developed the theory. JH and WL carried out the numerical modelling, and JH, PR and WL wrote the manuscript.

## 6 Funding

This work was supported by the National Natural Science Foundation of China (W2541021), Strategic Priority Research Program of the Chinese Academy of Science (XDB0650000), Shanghai Science and Technology Innovation Action Plan (21ZR1482700), Shanghai Science and Technology Plan Project Funding (23JC1410100) and Fuyang High-Level Talent Group Project.

## 7 Disclosures

The authors declare no conflicts of interest.

## 8 Data availability

Data underlying the results presented in this paper are not publicly available at this time but may be obtained from the authors upon reasonable request.

# Supplementary Information

## Oblique Bragg-Doppler scattering at relativistic grating created by backward Raman amplification in hydrogen

Wenhong Lai[1,2,3], Jiapeng Huang[1,*], Haozhe Guo[1], and Philip St.J. Russell[1]

[1]Russell Centre for Advanced Lightwave Science, Shanghai Institute of Optics and Fine Mechanics and Hangzhou Institute of Optics and Fine Mechanics, Chinese Academy of Sciences (CAS), Shanghai, China

[2]University of Chinese Academy of Sciences, Beijing, China

[3]Zhejiang Key Laboratory of Microstructured Specialty Optical Fiber, Hangzhou Institute of Optics and Fine Mechanics, Hangzhou, China

*Corresponding author: jiapenghuang@siom.ac.cn

## S1 Theory of Bragg-Doppler scattering

The efficiency of conversion between probe and Bragg scattered signal can be estimated using coupled mode theory[1], modified to include the Doppler shift at a rapidly moving grating. We write the electric field in the form:

$$E = a_{Pr}(x)e^{i(\vec{k}_{Pr}\cdot\vec{r}-\omega_{Pr}t)} + a_{Bs}(x)e^{i(\vec{k}_{Bs}\cdot\vec{r}-\omega_{Bs}t)} \tag{5}$$

where $a_{Pr}(x)$ and $a_{Bs}(x)$ are slowly-varying amplitudes of the probe and Bragg-scattered fields, $x$ is the coordinate perpendicular to the fibre axis in the plane containing the two beams, $\vec{k}_j$ are the wavevectors and $f_j = \omega_j/2\pi$ the frequencies. Defining a dephasing parameter $\vartheta = (\omega_{Bs}\cos\theta_{Bs} - \omega_{Pr}\cos\theta_{Pr})/c$ where $\theta_j$ is the angle between $\vec{k}_j$ and the $x$-axis (Fig. S1), and assuming negligible probe depletion, the conversion efficiency to the Bragg scattered signal takes the form:

$$\eta_{Bs} = \frac{\omega_{Bs}\cos\theta_{Bs}}{\omega_{Pr}\cos\theta_{Pr}}(\kappa_{Bs}h)^2\mathrm{sinc}^2\left(h\sqrt{\kappa_{Pr}\kappa_{Bs}+\vartheta^2/4}\right) \tag{6}$$

where $h$ is the effective width of the coherence wave, $\kappa_j = n_R\omega_j/(4c\cos\theta_j)$ are coupling constants, and $n_R$ is the index modulation of the coherence wave. For small deviations from the incident Bragg angle $\theta_{Pr}^0$, the dephasing parameter can be approximated as:

$$\vartheta \approx \frac{-4\pi f_{Pr}/c}{\sqrt{(1+(f_R/f_K)^2)((2f_{Pr}+f_R)^2/f_K^2-1)}}\Delta\theta_{Pr}, \qquad \Delta\theta_{Pr} \ll \pi \tag{7}$$

where $\Delta\theta_{Pr} = (\theta_{Pr} - \theta_{Pr}^0)$.

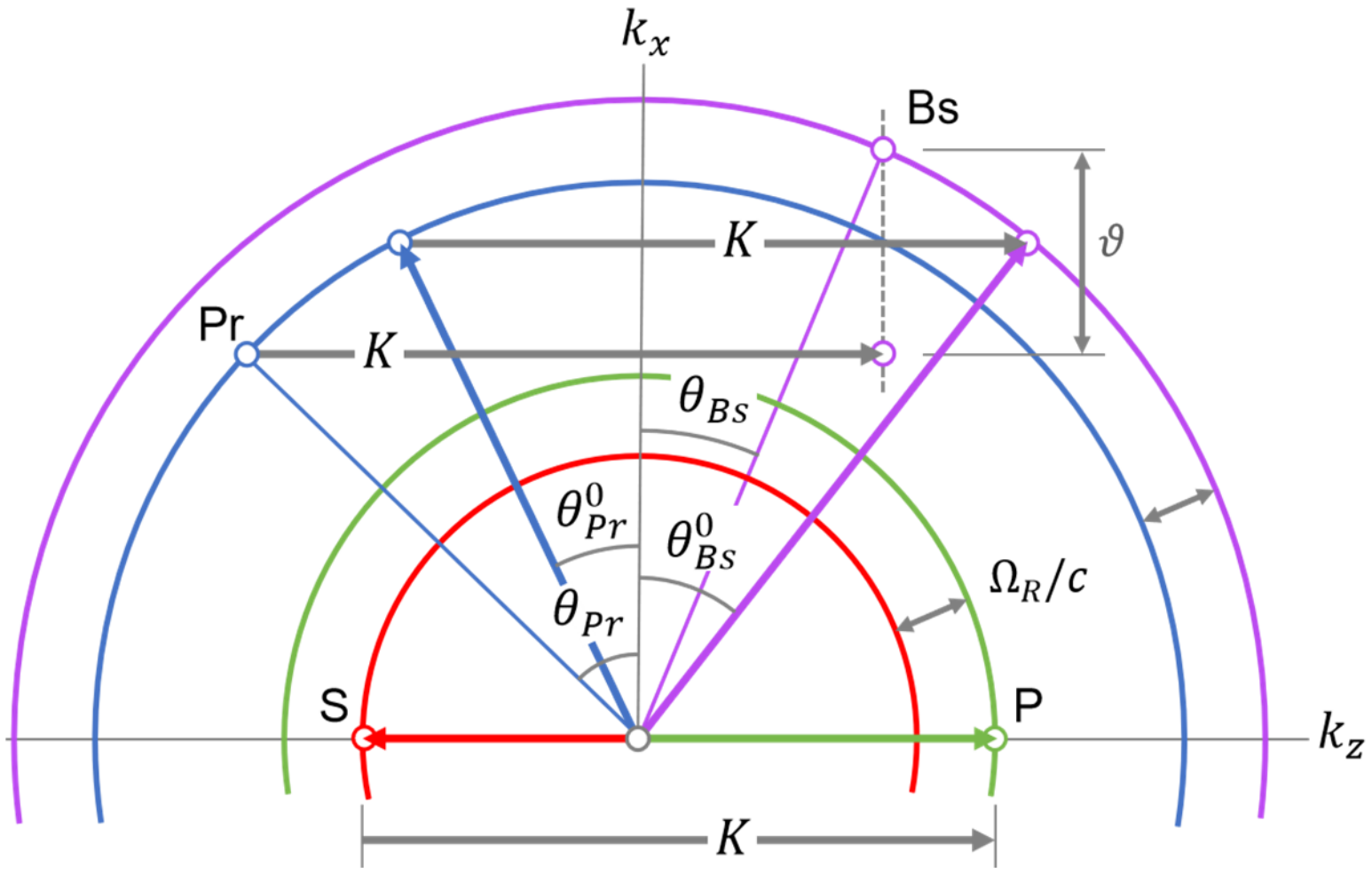

**Fig. S1**: Wavevector diagram for the case when the Bragg condition is violated by detuning the probe angle.

# S2 Spatio-temporal calculation

The full time-space plot is calculated by solving the Raman equations, following the procedure in[2]. The results, adjusted so as to optimize the coherence wave amplitude at the fibre input face, are shown in Fig. S2, along with experimental measurements made close to the fibre end-face.

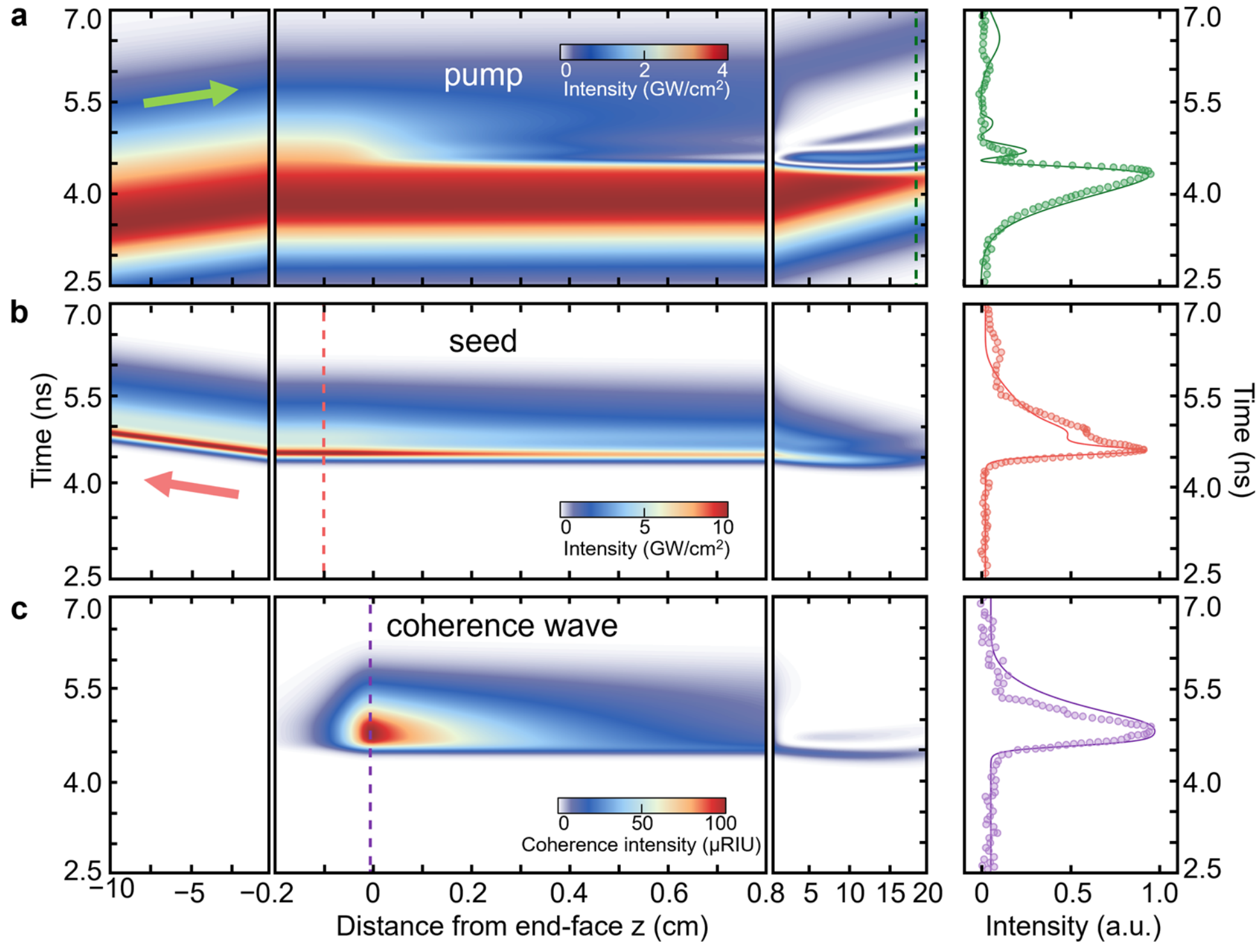

**Fig. S2**: Left: Simulated spatio-temporal evolution of (a) the pump, (b) backward Stokes signal and (c) coherence wave. Right: Corresponding temporal profiles at positions indicated by the dash lines.

# S3 Euler-Rodrigues rotation matrix

A convenient approach to resolving the components of a vector into a rotated coordinate system uses the Rodrigues matrix[3]:

$$[\mathbf{R}]_{\vec{u}}^{\psi} = [I] + [U]\sin\psi + [U]\cdot[U](1-\cos\psi) \tag{8}$$

where

$$[U] = \begin{bmatrix} 0 & -u_z & u_y \\ u_z & 0 & -u_x \\ -u_y & u_x & 0 \end{bmatrix}, \qquad \vec{u} = \begin{pmatrix} u_x \\ u_y \\ u_z \end{pmatrix} \tag{9}$$

and $\vec{u}$ is a unit vector defining the axis of rotation and $\psi$ is the rotation angle.